\documentclass[journal=jacsat,manuscript=article]{achemso}

\usepackage{chemformula} % Formula subscripts using \ch{}
\usepackage[T1]{fontenc} % Use modern font encodings
\author{Ange B. Chambissie Kameni}
\author{Alexandre Py-Renaudie}
\affiliation{CNRS Institut Photovoltaique d’Île de France (IPVF) UMR 9006, 18 Bvd Thomas Gobert , 91120 Palaiseau, France}
\author{Damien Garrot}
\affiliation{Université Paris-Saclay, UVSQ, CNRS, GEMaC, Versailles 78000, France}
\author{Baptiste Berenguier}
\affiliation{CNRS Institut Photovoltaique d’Île de France (IPVF) UMR 9006, 18 Bvd Thomas Gobert , 91120 Palaiseau, France}
\author{Guillaume Bouchez}
\affiliation{Université Paris-Saclay, UVSQ, CNRS, GEMaC, Versailles 78000, France}
\author{Davide Raffaele Ceratti}
\affiliation{ChimieParisTech-CNRS, PSL Research University, Institut de Recherche de Chimie Paris (IRCP), Paris, France Chimie ParisTech, France}
\author{Philip Schulz}
\affiliation{CNRS Institut Photovoltaique d’Île de France (IPVF) UMR 9006, 18 Bvd Thomas Gobert , 91120 Palaiseau, France}
\author{Jean-François Guillemoles}
\affiliation{CNRS Institut Photovoltaique d’Île de France (IPVF) UMR 9006, 18 Bvd Thomas Gobert , 91120 Palaiseau, France}
\author{Géraud Delport}
\affiliation{CNRS Institut Photovoltaique d’Île de France (IPVF) UMR 9006, 18 Bvd Thomas Gobert , 91120 Palaiseau, France}
\email{geraud.delport@cnrs.fr}

\title[An \textsf{achemso} demo]
  {Polar Indirect Valley as a Limiting Factor for Radiative Efficiency in Gold-Based Mixed-Valence Double Perovskites}

\abbreviations{IR,NMR,UV}

\begin{document}

\begin{tocentry}

%Some journals require a graphical entry for the Table of Contents.
%This should be laid out ``print ready'' so that the sizing of the

%Inside the \texttt{tocentry} environment, the font used is Helvetica
%8\,pt, as required by \emph{Journal of the American Chemical

%\includegraphics[scale = 1.16]{toc.png}

%The surrounding frame is 9\,cm by 3.5\,cm, which is the maximum
%permitted for \emph{Journal of the American Chemical Society}
%graphical table of content entries. The box will not resize if the
%content is too big: instead it will overflow the edge of the box.

%This box and the associated title will always be printed on a
%separate page at the end of the document.

\end{tocentry}

\begin{abstract}
 Double perovskites have emerged as promising alternatives to lead halide perovskites, aiming to mitigate challenges related to toxicity and chemical instability. Among them, mixed-valence gold halides such as $Cs_2Au^+Au^{3+}Cl_6$, which contain only a single type of metal cation in two oxidation states, stand out due to their unique structural and electronic properties. These materials exhibit strong absorption in the near-infrared range, making them attractive candidates for optoelectronic applications such as photovoltaics. In this work, we employ temperature-dependent optical spectroscopy techniques to demonstrate that these compounds exhibit particularly strong polar electron-phonon coupling, which has a profound impact on their optoelectronic properties. In particular, this coupling gives rise to a temperature-dependent absorption tail that reshapes the global spectral spectrum. We show that this tail leads to a forbidden band-egde recombination, which explains the reported difficulties in detecting a photoluminescence signal from this class of double perovskites.
\end{abstract}

\section{Introduction}
Lead-halide perovskites (LHPs) have attracted intense interest for their remarkable performance in light-emitting and light-harvesting applications, for instance leading to photovoltaic efficiencies exceeding 25$\%$ \cite{yang2023one,noman2024,green}. In response to concerns about lead toxicity and long-term stability, research efforts have increasingly focused on lead-free alternatives\cite{he2025homogeneous,hutchinson2023resolving,ruggeri2019controlling}. Among them, halide double perovskites (HDPs) have emerged as promising candidates\cite{volonakis2016lead,volonakis2017}. These compounds adopt the general formula \( \text{A}_2[\text{BX}_2][\text{B}'\text{X}_4] \), with A being a small cation, X being an halide and two distinct metallic cations occupying the B-sites. Despite this promise, the optoelectronic performances of HDPs have remained significantly inferior to that of LHPs, with photovoltaic efficiencies peaking around $6\%$\cite{ji2023challenges,zhang2022}. To explain this performance gap, recent studies have hypothesized that strong electron-phonon interactions in HDPs could lead to the rapid formation of small polarons\cite{Wright2021,tailor2022,ramesh2025}. In such a scenario, charge carriers become self-trapped within localized lattice distortions shortly after photoexcitation, impairing their transport.

However, the polaronic framework alone appears insufficient to fully account for the observed properties of many HDPs. In particular, HDPs exhibit reduced photoluminescence quantum yields (PLQYs) compared to their LHP counterparts\cite{li2021lead,Slavney2018,luo2018}. This trend is observed in excitonic compounds such as \( \text{Cs}_2\text{AgBiBr}_6 \) \cite{longo2020}, and is even more pronounced in non-excitonic ones \cite{Slavney2018}, which sometimes display no detectable photoluminescence. Since this effect appears to be generalized across various HDP compositions and irrespective of the crystallinity of the samples (thin films, single crystals, or nanocrystals), it cannot be convincingly explained by the presence of extrinsic defects. Therefore, an intrinsic origin must be sought to explain this phenomenon, which has dramatic consequences for potential applications. In this context, several studies have attributed these phenomena to the presence of parity-forbidden transitions between the conduction and valence band edges in HDPs~\cite{Slavney2018,luo2018}.

An alternative perspective on this debate is proposed here by investigating the potential role of electron-phonon interactions in the reduction of radiative recombination rates. To do so, the study was narrowed-down to a single HDP composition: \( \text{Cs}_2\text{Au}^+\text{Au}^{3+}\text{Cl}_6 \) (referred to as \textbf{AuClAu}), which presents a near-infrared bandgap that is promising for photovoltaic applications~\cite{pylow,Lindquist2023,Alexandre2024}. Moreover, no clear excitonic resonance is observed in its absorption spectrum~\cite{Lindquist2023}, although a low but measurable photoluminescence (PL) intensity has recently been detected~\cite{bajorowicz2020integrated,Alexandre2024,pylow}.

From a fundamental perspective, this compound presents the specificity that both \( \text{B} \) and \( \text{B}' \) sites are occupied by the same gold metal. This configuration is called a mixed-valence perovskite and offers specific optoelectronic effects due to their ability to distort upon internal or external stimulii\cite{pena2001chemical,nag2014oxide}. Indeed, a greater vibrational freedom is expected between the two sublattices with respect to conventional halide double perovskites (HDPs), where \( \text{B} \) and \( \text{B}' \) cations have fixed and well-separated ionic radii and valences. For this reason, \textbf{AuClAu} exhibits enhanced manifestations of the electron-phonon coupling, making it an ideal material for studying the consequences of such interactions. This is exemplified by recent results on effects such as transient polaronic distortions and intervalence charge transfer \cite{Wang2014,Kojima1994,ramesh2025,son2005photoinduced}.

\section{Results and discussion}

%\subsection{Outline}

\begin{figure*}
\includegraphics[scale = 0.9]{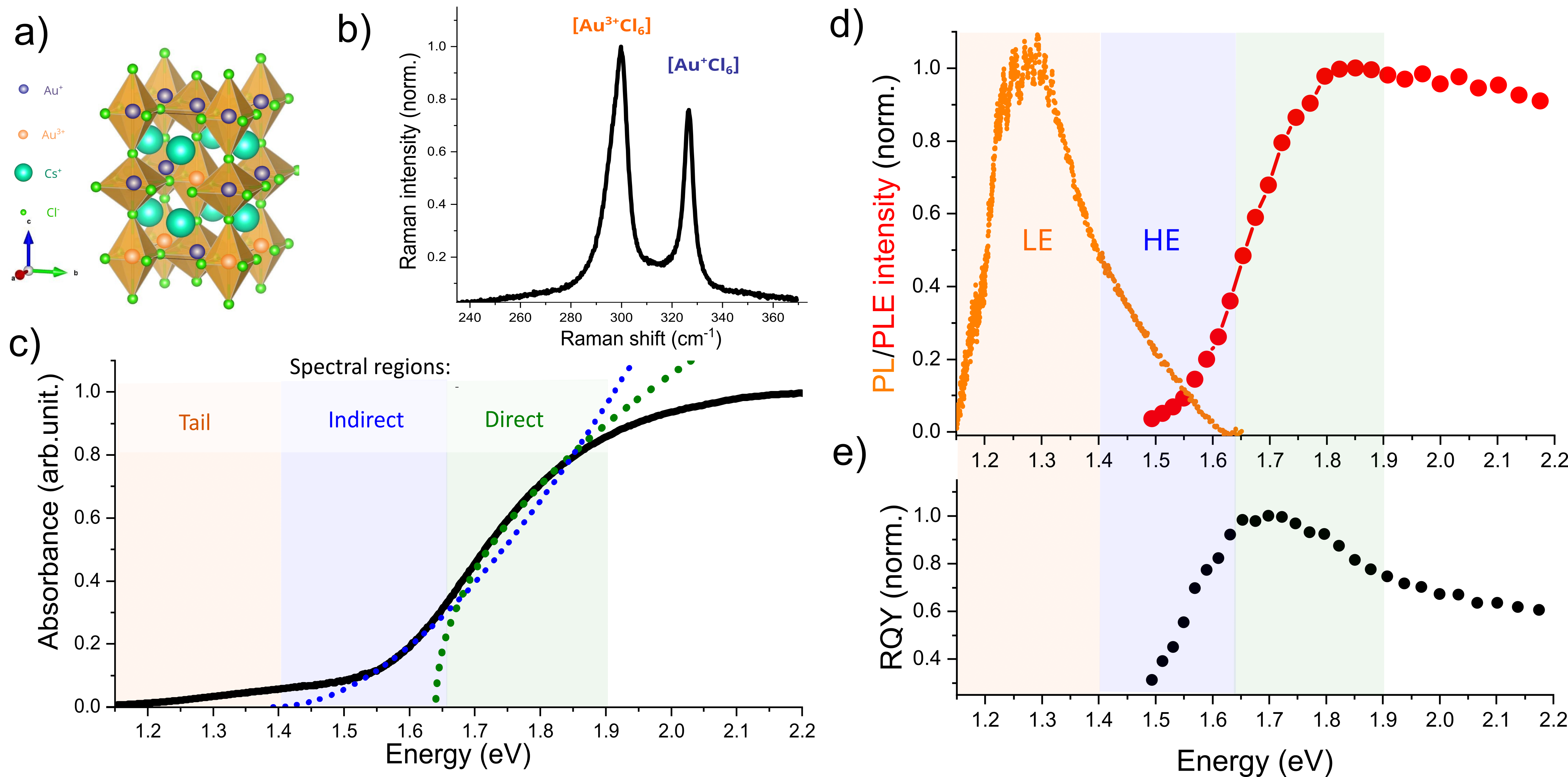}
\caption{\textbf{General properties of $\textbf{AuClAu}$} a) Crystallographic structure of the mixed-valence perovskite $Cs_2Au^+Au^{3+}Cl_6$. b) Raman spectra at 300K at 1.58 eV excitation exhibiting the main phonon B1g and A1g modes. c) Absorption spectrum at 300 K (in black), with two mathematical guidelines corresponding to the direct (green) and indirect (blue) band gap components, along with a sub-band tail (orange region) d) PL (orange) and PLE (red) spectra of $\textbf{AuClAu}$ at 300K. e) Relative quantum yield spectrum (RQY) extracted from the ratio of the PLE in d) and Absorption data in e). On panel c-e), the tail, indirect and direct regions are respectively highlighted in light orange,blue and green shades.}
\label{fig1}
\end{figure*}

%\bigskip

%{\large \textbf{Experimental results}}

%\bigskip 

In this paper, the impact of the electron-phonon coupling on carrier recombination channels in $\textbf{AuClAu}$ are investigated through a combination of Raman, absorption, photoluminescence (PL), photoluminescence excitation (PLE). Details on the sample preparation and experimental conditions can be found in the experimental section below.

To start, the Raman spectrum of $\textbf{AuClAu}$ upon pumping at 1.58 eV is detailed. As shown in Fig.\ref{fig1}b, such Raman spectrum of $\textbf{AuClAu}$ is dominated by two main resonances. These $A_{1g}$ and $B_{1g}$ modes are polar phonon modes associated to the symmetric and antisymmetric stretching vibrations within the equatorial plane of the [$AuCl_6$] octahedra, associated to a situation in which one octahedron fully expands while its first neighbours shrink accordingly. As seen in SI Section 1, each of these two peaks can be accurately fitted with a Lorentzian formula, centered respectively at 299 and 327 cm$^{-1}$. Based on previous studies\cite{RamanAu,liu1999}, the former region corresponds to a sum of $A_{1g}$ and $B_{1g}$ modes of the [Au$^{3+}$Cl$_6$] sublattice, while the latter is the $A_{1g}$ mode of the [Au$^+$Cl$_6$] sublattice. It is remarkable that $\textbf{AuClAu}$ possesses well-separated active modes for the specific elongation of the two [Au$^+$Cl$_6$] and [Au$^{3+}$Cl$_6$] sublattices. In other words, this spectrum demonstrates the ability of the two sublattices in $\textbf{AuClAu}$ to vibrate at separate frequencies.

In addition, Fig.\ref{fig1}c displays the absorption spectra of $\textbf{AuClAu}$, with an absorption intensity that progressively decreases between 2.2 and 1.2 eV. By analogy with conventional semiconductors, the high-energy region (above 1.9 eV), where the decay slope is relatively small, can be attributed to the free carrier continuum or the absorption onset (as it will be referred in the following discussion). Consequently, the lower-energy region with a steeper decay slope (between 1.4 and 1.9 eV) corresponds to the band-edge. The shape of this region is highly complex and cannot be adequately fitted using a simple Tauc plot approach (see Fig.~\ref{fig1}c and SI Section 2 for details). Nevertheless, tentative estimations of direct (green guideline) and indirect (blue guideline) components can be made using separate Tauc plots. The positions of these two bandgaps are $E_{\text{indirect}}=1.4$ eV and $E_{\text{direct}}=1.63$ eV, respectively. Below the fitted indirect bandgap region (between 1.4 and 1.2 eV), the absorption spectrum still exhibits a low-energy tail (see Fig.~\ref{fig1}c and SI Section 2). A similarly complex band-edge structure appears to be a general tendency in the family of HDPs \cite{Steele2018,Slavney2016}, which typically exhibit large indirect and tail absorption components. However, in this case, no excitonic resonance can be detected, in contrast to other HDPs that present strong excitonic resonances even in the three-dimensional (and therefore non-confined) crystallographic structure.

  \begin{figure*}
\includegraphics[scale = 0.4]{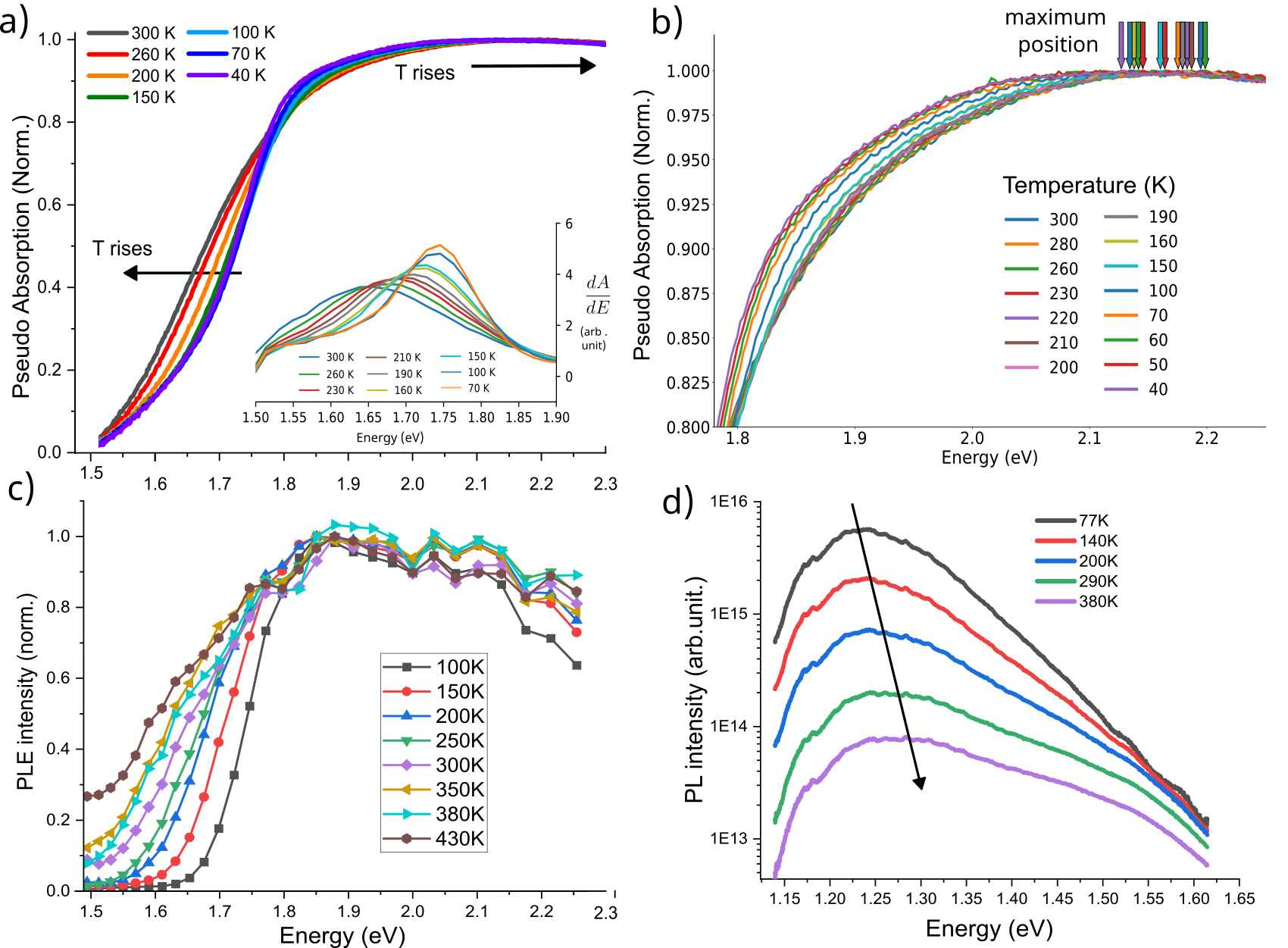}
\caption{{\textbf{Evolution with temperature of the electronic properties.} (a) Evolution of the pseudo-absorption spectra extracted from temperature-dependent transmission spectroscopy (see SI Section 5 for details on the extraction method). Inset: evolution of the derivative of the pseudo-absorption spectra. For each temperature, the peak value of this derivative indicates the energy corresponding to the steepest slope across the band-edge region. (b) Zoom on the high-energy absorption onset region to highlight the position of maximum absorption and its blueshift with increasing temperature. (c) Evolution of the PLE spectra as a function of temperature. (d) Evolution of the PL spectra as a function of temperature. These spectra have been calibrated to account for the spectra response of the optical setup and correspond to an excitation at 2.33 eV. The black arrow highlights the progressive blueshift of the central wavelength with increasing temperature.}}
\label{fig2}
\end{figure*}

%highlighting the two low energy (e) Evolution of the spectrally integrated PL intensities of the HE (blue) and LE (orange) resonances as a function of temperature along with the Boltzmann equilibrium fitting function used to extract the free activation energies G$_1$ and G$_2$.f) Schematic of the possible band structure of $\textbf{AuClAu}$, exhibiting the static direct gap, the indirect gap and the polar indirect tail with the implication on the radiative and non-radiative recombination pathways. The two Free activation energies G$_1$ and G$_2$ are represented in red as saddle points in the conduction band structure. Here the indirect tail has been displayed in the conduction band (CB) but could also be present in the valence band (VB).}}

% and can be separated in two components (see Fig.\ref{fig1}d) and Fig.\ref{fig2}d)). The first one is a weak high-energy tail (HE) that is located close to the indirect bandgap region between 1.4 and 1.6 eV. Therefore, this component can be assigned to the radiative recombination of carriers from indirect region of the band diagram. A second and more intense PL component is located 

 To provide another perspective on these effects, the PL and PLE spectra at 300K are displayed in Fig.\ref{fig1}d. The PL spectrum is broad and centered at 1.27 eV, much below the indirect gap region. In addition, the width of this resonance is estimated to 200 meV, which is much larger than $k_BT$. In addition, the measured PLQY here is around 1E-6 under 2.2 eV excitation (see SI Section 3 for details). As for the PLE spectrum, its general shape is quite different from the absorption one (see Fig. \ref{fig1} c-d). It exhibits a maximum intensity around 1.82 eV—within the direct bandgap region—and gradually decreases in the indirect bandgap region down to 1.49 eV. Interestingly, this decrease is more abrupt compared to the absorption spectra.

Additional information on these complex electronic features can be obtained by analysing the temperature-dependent (pseudo) absorption spectra (estimated here only from transmission spectra, see Fig.\ref{fig2}a) and SI Section 4 for details). A significant redshift between 40 K and 300 K, is observable at the band-edge. Along with the redshift with temperature, an decrease of the band-edge slope is observed as the temperature increases. These effects are better visualised by monitoring the maximum value and the energy position of the pseudo-absorption derivative (see Fig.\ref{fig2}a inset). In contrast with the redshift of the band-edge, one can observe a blueshift of the upper part of the absorption onset (see Fig.\ref{fig2}b). Such onset position is here defined as the energy position at which the absorption is maximum and is located in the 2.0-2.3 eV range.

To complete this analysis of the temperature-dependent band-edge, the evolution of the PLE spectra is highlighted in Fig.\ref{fig2}b). Within the same temperature range, the slope of the PLE tail decreases, in a proportion that seems much more pronounced than in absorption. Finally, this overview can be completed by analysing the evolution of the PL spectral and temporal decays for different temperatures. As seen in Fig. \ref{fig2}d), the PL intensity decreases of almost two orders of magnitude between 77 and 350K. Interestingly, the amplitude of this decrease is much larger than the one usually reported in LHPs\cite{zheng2017temperature, gauthron2010optical}. In addition, the PL decay are also affected by temperature changes. As seen in SI section 5, the PL transient decay at room temperature is short—very close to the instrument response function (IRF)—making it difficult to study with high accuracy. However, as the temperature decreases, a doubling of the associated lifetime $\tau_{PL}$ is observed, with an extracted lifetime extending from $250~ps$ at 300 K to $525$ $ps$ at 77K (see SI Section 7 for details). At 77 K temperature, the PL intensity already decays over four orders of magnitudes within 8 ns. In addition, a much longer decay component is still observable over the 100 $ns$ - 10 $\mu s$ timescale (see SI Section 5). This indicates that at this low temperature, some carriers survive much longer than at room temperature. 

%\bigskip

%{\large \textbf{Thermal splitting of the absorption spectrum}}

\bigskip 
 
The first striking effect that emerges from the results in Fig. \ref{fig1} is the complexity of the band-edge, which not only exhibits a direct bandgap region but also a lower-energy substructure with an indirect character. Based on the results from Fig.\ref{fig2}, it appears that the energy splitting between these two regions increases with temperature. This distinction suggests that different physical effects impact these regions, as further analysed in the following discussion.

\begin{figure*}
\includegraphics[scale = 0.55]{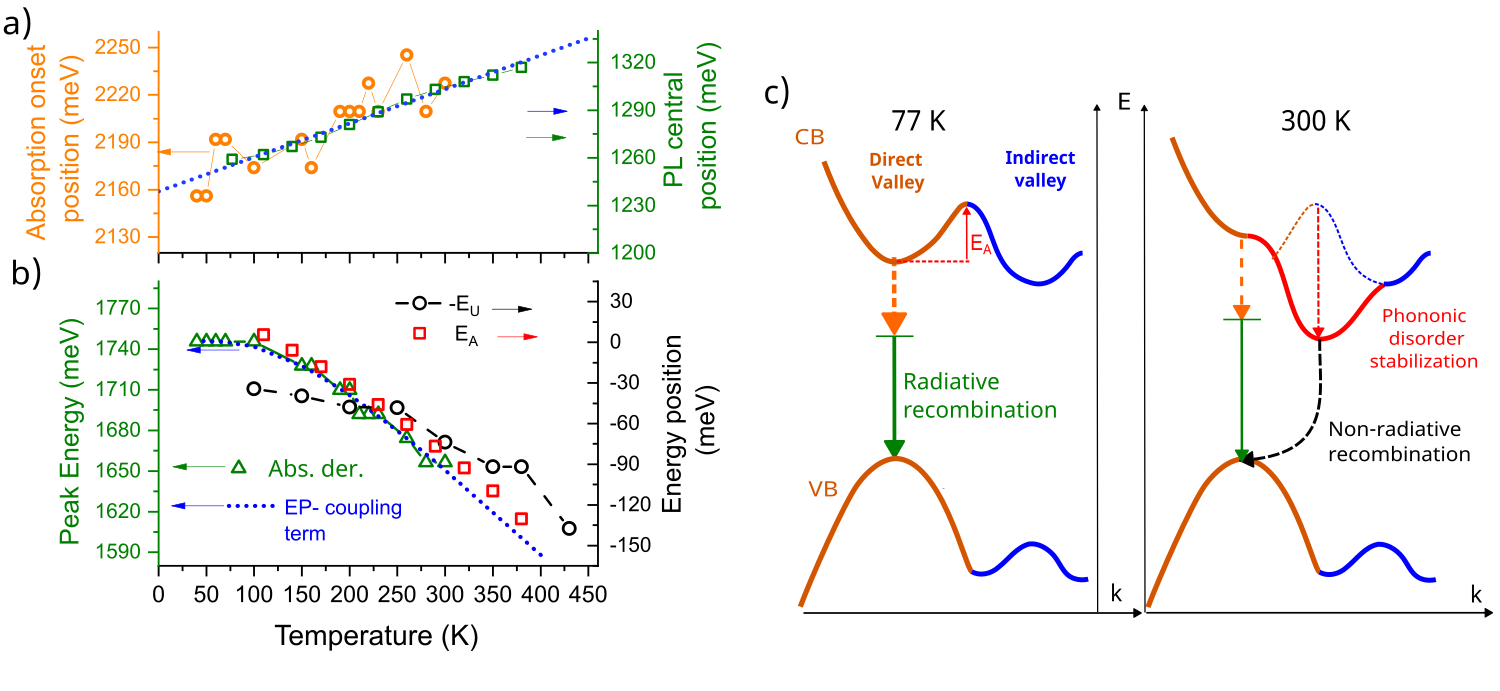}
\caption{{\textbf{Quantitative analysis of of the blueshift/redshift splitting} 
a) Overlay on the same graph of the two energies that blueshift with temperature: the high-energy absorption onset (orange) and the central position of the PL resonance (green). The blue dotted line corresponds to the linear thermal expansion blueshift term, which has been used to fit the absorption onset, as explained in the main text. b) Overlay on the same graph of all the energy values that redshift with temperature: the maximum position of the pseudo-absorption derivative (green), the Urbach energy (shown here in black as $-E_U$), and the extracted saddle-point energy (red). These curves have been offset so that the zero in Urbach/activation energy approximately corresponds to the band-edge derivative position in the low-temperature limit (where no polar phonons are present in the system). The blue dotted line corresponds to the electron-phonon coupling redshift model, which has been used to fit the absorption derivative position, as explained in the main text. c) Schematic of the possible band structure of $\textbf{AuClAu}$, showing the static direct gap and the indirect gap region. Between 77 K and 300 K, the main difference lies in the lowering of the saddle-point activation energy $E_A$, which controls the direct-to-indirect valley crossover. Here, the saddle point has been depicted in the conduction band (CB), but it could equally be present in the valence band (VB) without changing the described process.
}}
\label{fig2-b}
\end{figure*}

Firstly, the high-energy direct part of the absorption spectrum experiences a 70 meV blueshift between 40 and 300 K. This effect is mostly visible in the free carrier continuum part around 2.0–2.3 eV (see Fig.\ref{fig2}a and \ref{fig2-b}a). Such an effect is common in many LHPs and is associated with the thermal expansion of the crystal as the temperature increases\cite{yu2021,Baldwin2021}.Based on literature, such blueshift tendency can be fitted with the usual law\cite{yu2021} :

 \begin{equation} \label{TE}
 E_{onset}(T)= E_{onset}(0)+A_{TE}\cdot T \end{equation}

This allows for the evaluation of the optical thermal expansion coefficient, which is found to be $A_{TE} = 0.21$ meV.K$^{-1}$. Interestingly, this value is approximately four times larger than the value reported for CsPbBr$_3$ \cite{Yuan2021} and three times larger than that for CsPbCl$_3$ \cite{Handa2020}. While a extensive band structure modelling would be necessary to fully understand such effect, such a high $A_{TE}$ value could originate from a high thermal expansion coefficient, that directly probes the changes in lattice parameters (through temperature-dependent X-Ray measurements). However, as seen in Fig. S12-13, such value for $\textbf{AuClAu}$ is $3.3 \times 10^{-5}$ K$^{-1}$. Such value is close to the one reported for MAPbCl$_3$ of $5.5 \times 10^{-5}$ K$^{-1}$ \cite{Handa2020}. Therefore, this mechanical effect cannot simply explain the large optical $A_{TE}$ coefficient. Interestingly, this particularly large thermal expansion coefficient can be related to the softness of the perovskite crystal, a property that was investigated in the context of a previous study on the self-healing properties of $\textbf{AuClAu}$ \cite{Alexandre2024}.

As seen in Fig.~\ref{fig2}c and \ref{fig2-b}a, the extracted central position of the PL resonance also exhibits a blueshift with temperature. While the PL resonance is positioned approximately 900 meV below the absorption onset, this offset appears to be temperature-independent, and the two energies seem to follow the same trend across the temperature range. This similar trend suggests a common origin for the two effects. In conventional LHPs (e.g., MAPbI$_3$ \cite{diab2016narrow}, CsPbBr$_3$ \cite{Yuan2021}, etc.), where the Stokes shift is minimal, the PL radiative emission can be seen as a simple band-to-band transition. As a consequence, the temperature shift of the PL resonance should be the same as that of the absorption band-edge. However, the situation is quite different in \textbf{AuClAu}, in which the PL resonance reported above is highly Stokes-shifted and therefore less related to the band-edge electronic states. In other HDPs, such a large Stokes shift is frequently attributed to small-polaron formation, in which the photocarriers become self-trapped due to their stabilizing interaction with surrounding phonons\cite{Steele2018,Hu2016,Wu2019}. As phonons are strongly involved in such polaronic distortion, one could expect to lose the strong correlation between the absorption onset electronic states and the polaronic states from which photons are eventually emitted. It is therefore quite remarkable to observe that such a common blueshift is maintained through the polaronic distortion, suggesting that this correlation is preserved through the polaronic distortion. In particular, this effect suggests that the polaronic stabilization energy remains the same across the temperature range, while the observed shift is solely due to the blueshift of the non-distorted direct region absorption. Additionally, this fact reinforces the idea that the radiative emission from polaronic states mainly originates from states in the direct part of the absorption spectrum.

In contrast to the high-energy blueshift, the low-energy part of the absorption band-edge exhibits a progressive redshift with increasing temperature. This effect is better visualized in the inset of Fig. \ref{fig2}a, where the derivative of the absorption onset in the 1.5 to 1.9 eV region is plotted as a function of temperature. These curves allow tracking of the peak derivative energy, corresponding to the point of highest slope of the band-edge. Thanks to this method, the redshift can be evaluated to -90 meV between 40 and 300 K, as seen in Fig. \ref{fig2-b}b. As seen in Fig. \ref{fig2-b}b, this redshift can be accurately fitted with the conventional law \cite{yu2021}:

\begin{equation}\label{EP}
E_{onset}(T) = E_{onset}(0) + A_{EP} \cdot \frac{2}{e^{E_{ph}/(k_B T)} - 1}
\end{equation}

where $A_{EP}$ is the electron-phonon coupling term and $E_{ph}$ is the associated phonon energy. This fit yields the values $A_{EP} = -153$ meV and $E_{ph} = 37$ meV. Interestingly, this last value matches exactly the energy of the polar phonon mode A$_{1g}$ mode evidenced from the previous Raman spectrum displayed in Fig. \ref{fig1}b. On the other hand, the $A_{EP}$ is close to the previously reported value for CsPbBr$_3$ \cite{Yuan2021}.

While a blueshift with temperature is expected for perovskites close to the three-dimensional cubic structure, a redshift of the absorption onset has already been reported in several low-dimensional structures \cite{wolf,Baldwin2021,wang2019temperature}. This effect has been attributed to an enhanced electron-phonon coupling in such distorted crystallographic structures, which amplifies the effect of the electron-phonon coupling. However, in many LHPs, the fitting of the temperature-dependent absorption edge requires a combination of the blueshifting thermal expansion term from \eqref{TE} and the redshifting electron-phonon coupling from \eqref{EP}. However, the situation reported here $\textbf{AuClAu}$ is quite different, as it appears that these two competing effects have an impact on separate energy regions. As a result, they effectively lead to an energy splitting within the absorption spectrum.

%\bigskip 

%{\large \textbf{Forbidden radiative recombination}}

%\bigskip 

To go analyse further this electron-phonon coupling effect, Yu et al. have observed that such redshift in CsPbBr$_3$ nanoplatelets\cite{yu2021} was correlated to the presence of a low-energy absorption tail. Interestingly, these results present similarities to the situation described above for $\textbf{AuClAu}$. This low-energy tail can be related to the concept of Urbach tails \cite{Studenyak2014}, used to describe the present of low-energy states that are able to locally trap carriers. Such energetic disorder can originate from local heterogeneities\cite{klick1964,Peretti} (e.g., variations in composition or strain). However, in polar semiconductors such as LHPs, these shallow levels are predominantly linked to phonon-induced electronic disorder. Consequently, these effects tend to vanish for a temperature low enough to significantly decrease the concentration of polar phonons \cite{Zhang2023,Falsini2022,Studenyak2014}.

Interestingly, this low-energy tail scenario could nicely explain the absorption redshift observed in \textbf{AuClAu}. However, there is a significant difference between the situation of \textbf{AuClAu} and the one described by Yu et al. for CsPbBr$_3$ nanoplatelets\cite{yu2021}. Indeed, for the nanoplatelets, the PL peak energy closely follows the band-edge position (e.g. it is always located at the top of the low-energy tail). In \textbf{AuClAu}, the band-edge emission is negligible, while the only detectable PL signal is highly Stokes-shifted. Moreover, the PL peak position follows the blueshift of the absorption onset rather than the redshift of the band-egde. 

As the PL spectra in \textbf{AuClAu} follow the blueshift of the high-energy absorption onset, one can hypothesize that the low emission signal originates from this specific energy region. Thus, the high-energy onset can be regarded as (at least partially) allowed. On the contrary, the emission from the lower energy band-edge appears quenched, which could be the signature of a forbidden transition, similar to previously reported results in other HDPs \cite{Slavney2018,luo2018}. 

To investigate this forbidden character and its connection to the low-energy absorption region, it is instructive to examine the PLE spectra. As mentioned earlier, the PLE intensity of \textbf{AuClAu} decreases more abruptly than the absorption intensity upon excitation below the direct region. One way to visualise this drop in PL intensity is to construct the relative quantum yield (RQY) value for each excitation energy $E$ by dividing the PLE intensity $I_{PLE}$ with the previously measured absorption coefficient $A(E)$:

\begin{equation} I_{PLE}(E) = \gamma \times \text{PLQY}(E) \times A(E) \end{equation}

Here, it has been assumed that PLE and absorption intensities are proportional to each other via a constant $\gamma$ and the PLQY at a given excitation energy $E$(see SI Section 3 for details). A constant excitation power was applied at each energy to prevent variations in PLQY due to power-dependent recombination effects, such as trap filling. \cite{delport,Baldwin2021}. As seen in Fig. \ref{fig1}e, the RQY remains relatively high within the high-energy direct region but decreases below 1.64 eV, reaching only 30$\%$ of its maximal value at 1.49 eV. Since it quantifies the ratio between emitted and incident photons, the RQY directly probes the competition between radiative and non-radiative recombination pathways. Thus, the probability of radiative recombination appears to be maximal when carriers are initially injected within the direct energy region. On the other hand, as the injection energy decreases below $E_{\text{direct}}$, the RQY starts to decrease. This points towards a larger proportion of non-radiative events that validates the hypothesis of the low-energy-tail having an amplified forbidden character. 

At this point, a question arises regarding the role of electron-phonon coupling in this forbidden character. Indeed, given that such an effect is responsible for the low-energy tail in many LHPs, it might also be the cause of the associated forbidden character in \textbf{AuClAu}. To validate this hypothesis, it is insightful to analyse the temperature evolution of the PLE spectrum. As seen in Fig. \ref{fig2}c, the PLE intensity in the indirect region increases with temperature. Given that the PLE is directly affected by PLQY changes, it makes sense that the growth of this indirect tail appears much larger in PLE than in the absorption spectra from Fig.\ref{fig2}a. Using Urbach tail models\cite{Wu2019,Zhang2023, Peretti}, it is possible to extract the associated energy $E_U$. This energy represents the average energy position of the phonon-induced trap levels with respect to the absorption band-edge. Therefore, the value of -E$_U$ can be readily compared to the relative redshift of the band-edge position (see Fig. \ref{fig2-b}b). 

For each temperature, E$_U$ was obtained by fitting the exponentially decreasing part of the PLE tail (see SI Section 7). As seen in Fig. \ref{fig2-b}b, the obtained $E_U$ increases from 30 meV at 100 K to around 140 meV at 430 K. In the low-temperature limit, $E_U$ should converge towards the static Urbach energy induced by temperature-independent disorder effects, such as compositional heterogeneities \cite{Studenyak2014}. The obtained value of 30 meV at 100 K might constitute an upper bound for this static $E_U$ value. Above 200 K, one can observe that the slope of the Urbach energy is very close to that of the redshift of the band-edge due to the growth of the indirect tail. This can be understood by stating that the polar disorder is so strong in \textbf{AuClAu} that it not only leads to a growth of the Urbach energy but also results in a massive redshift of a whole part of the absorption band-edge. Another way to quantify the massive impact of such polar distortion is to compared the growth rate of $E_U$ with temperature. In the asymptotic limit\cite{Yamada2022,Zhang2023}, the value of $E_U$ is expected to grow according to the affine law $E_U \simeq a + b*T$ with $b=k_B=8.62\cdot10^{-2}$ meV.$K^{-1}$. However for $\textbf{AuClAu}$, the fitting parameters are $a=-66$ meV and $b=4.5\cdot10^{-1}$ meV.$K^{-1}$. Therefore, the slope at which the Urbach energy grows with temperature is approximately 5 times larger than the expected value. This growth rate explains why the indirect polar trapping is so efficient in $\textbf{AuClAu}$ at room temperature.

At this point, it is insightful to compare the redshift of the absorption band-edge and the growth of $E_U$. As displayed in Fig. \ref{fig2-b}b, the two values follow a similar trend with temperature, which tends to prove that they originate from the same electron-phonon phenomenon. To be more quantitative, the evolution of $E_U$ can also be fitted with the equation \ref{EP}, yielding a value of 42 meV. This number is slightly larger but comparable to the value obtained from the absorption band edge. However, it is still within the range of the observed polar modes in the Raman spectrum (see Fig. \ref{fig1}b).

All of these elements seem to confirm the presence of a common electron-phonon coupling phenomenon that both generates a low-energy tail, and a suppressed band-edge radiative recombination in \textbf{AuClAu}. Interestingly, this approach shows similarities with the work of Wu et al.\cite{Wu2019}, who reported the presence of an indirect tail in other LHPs. They attributed this effect to polar-induced splitting of the electronic bands into separate sub-bands or valleys. As these valleys possess different momentum and/or spin characteristics, vertical band-to-band recombination would become forbidden, while radiative recombination would require the mediation of a phonon. In such a framework, the polar-induced states are no longer seen as a distribution of isolated trap states but rather as a completely new electronic valley having a strong impact on carrier dynamics.
 
%Beyond the conventional indirect character, it has been shown that the absorption within the indirect regions decreases at temperatures at which the population of optical phonons becomes negligible. Such phenomenon nicely fits the formalism of phonon-induced indirect bands\cite{Wu2019,Zhang2023}. More precisely, the slope of the tail observed in PLE can be accurately fitted with an exponential increase with energy. For each temperature this fitting allows to extract the associated Urbach energy $E_U$. At 100K, $E_U$ is close to a value of 30 meV which can be interpreted as the static Urbach energy attributed to temperature-independent disorder effects (such as compositional heterogeneities \cite{Studenyak2014}). At larger temperature, the Urbach energy progressively increases as the phonon concentration increases. At large enough temperatures, this curve should follow a*k$_B$T asymptotic law\cite{Yamada2022,Zhang2023}. Here, the data seem to follow this law only for intermediate temperatures (100-250 K) but then a even larger slope is observed and remain to explained (see Fig. \ref{fig2-b}d).

%\bigskip 

%{\large \textbf{Saddle-point driving the recombination}}
%\bigskip 

Within the framework of this indirect tail scenario, it is not clear how it would quantitatively impact the Radiative emission properties. As seen in Fig. \ref{fig2}c-d and SI, both PL intensities and lifetimes decrease significantly with temperature. This effect can be understood by considering a temperature dependent competition between the different recombination channels. In the following, a simple formalism will be developed to describe such effect. It will be based on the assumption that only two recombination channels exist in \textbf{AuClAu}. The first is the radiative recombination of redshifted self-trapped carriers, which is favoured when carriers primarily reach the direct energy gap region. The second is a non-radiative process that occurs when carriers fall into the polar indirect tail region.

\begin{figure*}
\includegraphics[scale = 0.42]{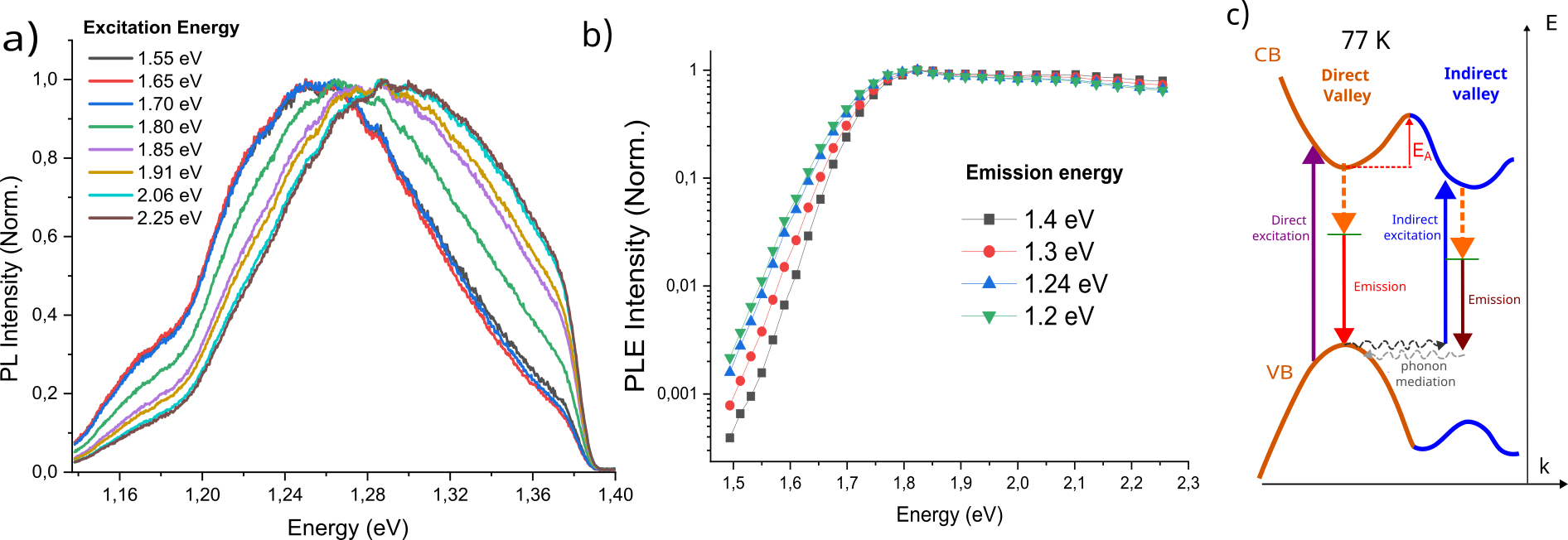}
\caption{{\textbf{Influence of the Saddle point at low temperature} 
 a) Dependence of the PL spectra at 77K on the selected excitation energy. PL spectra have been PL spectrally calibrated to account for the spectral response of the optical setup. b) Dependence of the excitation (PLE) spectra at 77K on the selected energy within the PL emission spectrum. c) Schematic of the two radiative recombination pathways coexisting at 77K : the direct excitation (purple) leading to blueshifted self-trapped PL and the indirect excitation (blue) leading to redshifted self-trapped PL. 
}}
\label{fig4}
\end{figure*}
 
 The rates associated with these two phenomena will be labelled as $k_{R}$ (radiative) for the self-trapped emission and $k_{NR}$ (non-radiative) for the trapping of carriers due to the polar distortion. In such a scenario, the competition between $k_{NR}$ and $k_{R}$ can also be expressed in terms of an activation energy $E_A$. As seen in Fig. \ref{fig2-b}c, this activation energy can be viewed as a saddle point in the band structure. As a result, carriers that lack the sufficient energy to overcome this saddle point will remain in the direct part and are more likely to recombine radiatively. For simplicity, it will be assumed that this energy saddle point is in the conduction band, although its presence in the valence band instead would not alter the underlying mechanism. Based on the corresponding Arrhenius law, the PL intensity $I_{PL}(T)$ should follow the law :

 \begin{equation} \label{pl} I_{PL}(T)= I_{PL}(0)\dfrac{1}{1+\frac{k_{NR}(T)}{k_{R}}} = I_{PL}(0)\dfrac{1}{1+e^{-\frac{E_A}{k_B T}}} \end{equation}
 
Unfortunately, such a simple law fails to properly fit the PL intensities. Indeed, a static activation energy model cannot account for such a significant drop in the PL intensity. However, this tendency could be consistent with a temperature-dependent activation energy. Therefore, an alternative model is developed based on the preceding discussion on polar disorder. The previous discussion has evidenced a lowering of the band-edge position, which is, in principle, close to the saddle point energy position. Therefore, it appears reasonable to assume that the saddle point energy also experiences a downshift of comparable amplitude. To evaluate the temperature evolution of such activation energy, equation \eqref{pl} has been reversed :

 \begin{equation} \label{pl} E_A(T)= E_A(0)-k_B T\cdot ln\left( \frac{I_{PL}(0)}{I_{PL}(T)}-1)\right) \end{equation}

The values of $E_A(T)$ are then displayed in Fig. \ref{fig2-b} b) for comparison. They evolve from a positive value of 15 meV at $T = 100~\mathrm{K}$ to approximately -130 meV around 300 K. Interestingly, this curve follows the same trends as that of the absorption band-edge and $E_U$. Such a result tends to validate the hypothesis that polar-induced indirect tail is also responsible for the lowering of the saddle point energy. One interesting aspect is that the values appear to follow an affine tendency with temperature. Therefore, this law resembles that of free energy $G = H - T \cdot S$. This observation is appealing in the context of activation energy and Arrhenius laws, in which the existing entropy term is too often neglected, leading to poorly fitting functions. Interestingly, non-Arrhenius laws can be found when kinetic phenomena are associated with an increase in disorder\cite{bixon}. A comprehensive thermodynamic description of the polar distortion is beyond the scope of this study. However, including an entropic term in the free-energy barrier appears appropriate for the situation described above. Indeed, electron-phonon interactions shift the absorbing states from high-energy and well-defined positions at low temperature toward lower-energy and more disordered states at higher temperature.

Thanks to this free activation energy formalism, a global scenario emerges to explain every spectroscopic features detailed above. It is based on the fact that the main element that changes with temperature is the energetic position of this saddle point that exists between, the direct and the indirect part of the band structure (see Fig. \ref{fig2-b} c). As the energy of this saddle point decreases, the direct-to-indirect carrier transfer becomes increasingly favoured. Consequently, at room temperature, the saddle point is suppressed, and the majority of carriers transition freely from the direct to the indirect region, accounting for the very low PL efficiency evidenced at room temperature in \textbf{AuClAu}. In parrallel, the decrease of this saddle point position also explain the redshift of the Urbach energy and of the absorption band-edge.

On the other hand, at low temperature, the Saddle point energy is still high enough to prevent the efficient transfer of carrier between the direct and the indirect valleys. This effect can be seen by analysing how the position of the PL spectra varies depending on the excitation energy. The central position of the PL spectra progressively decreases as the excitation energy decreases from the high-energy onset (at 2.25--1.9 eV, see Fig. \ref{fig4}a) of the free carrier continuum toward the indirect tail (at 1.7 to 1.55 eV). At 77K, carriers that are injected in the indirect region will statistically recombine at lower energies than the one injected in the direct region. To complete this analysis, it is possible to measure the PLE spectra at 77K corresponding to different emission energies, either at highest or at the lower part of the broad PL spectra. As seen in Fig. \ref{fig4}b), the amplitude of the indirect tail in these PLE spectra strongly depends on the emission spectra. While such indirect tail is minimal when the PL collected at high energy (1.4 eV) it progressively grows as the selected PL energy is set to lower values. As displayed in Fig. \ref{fig4}c), this feature demonstrates that the excitation in the indirect tail would lead to a lower emission than the excitation in the direct part of the spectrum. Interestingly, such separation is not observed at room temperature, at which the position of the PL peak does not depend on the excitation energy (see Fig. S17). This distinction can be explained by the temperature-dependent amplitude of the saddle point that differs strongly at the two temperature. It also illustrates the role of this saddle point energy in the scattering and redistribution of carriers between the direct and the indirect regions.

\bigskip

{\large \textbf{Conclusion}}

\bigskip 
In brief, this study provides new insights into the complexity and richness of electron-phonon interaction in gold-based double perovskites thanks to an innovative combination of spectroscopic techniques. Evidences have been presented that this interaction splits the energetic diagram in two regions. The high-energy free-carrier continuum appears unaffected by the distortion, as it exhibits the conventional blueshift with increasing temperature. In contrast, the low-energy band edge is highly sensitive to the polar distortion, as evidenced by the emergence of a low-energy absorption tail. The growth of this indirect tail is accompanied by a decrease of the luminescence efficiency. Such indirect character is mostly present at high temperature, as it requires the polar distortion to reduce the activation energy of the saddle point. This study paves the way for a deeper understanding of the influence of polar interactions on the optoelectronic properties of such compounds. It has drawn parallels with other double and single perovskite materials to provide a comprehensive understanding of the impact of polar tails. In addition, this report draws attention to the necessity of understanding and minimizing the impact of such tails on carrier transport and non-radiative recombination in lead-halide perovskites.

\section{Experimental}

The analysis of the absorption and PLE spectra detailed above, covering several orders of magnitude of absorptivity, is here made possible by the use of a "solid dilution" approach, where a $\textbf{AuClAu}$ microscopic powder (grown by a method previously published \cite{pylow}) is embedded in a CsCl pellet, serving as an inert and quasi-transparent matrix. This simple technique enables the precise identification of energy resonances while mitigating common artefacts (see details in SI section 8). \\

\textbf{Chemicals} -- Cesium chloride (ultra-dry, 99.9\% metal basis, Thermo Scientific Chemicals), gold(I) chloride (99.9\% metal basis, Thermo Scientific Chemicals), and hydrogen tetrachloroaurate hydrate (III) (99.9\% metal basis, Au 49\% min, Thermo Scientific Chemicals). A solution of CsCl in HCl at a concentration of 4 $\times$ 10$^{-2}$ mol$\cdot$L$^{-1}$ was added to a solution of AuCl and AuCl$_3$ in HCl at concentrations of 1 $\times$ 10$^{-2}$ mol$\cdot$L$^{-1}$. The molar ratio is thus 2:1 (n$_\text{CsCl}$ = 2 $\times$ (n$_\text{AuCl}$ + n$_\text{AuCl$_3$}$)).\\

\textbf{Synthesis} -- The addition of the above solution was performed under ambient conditions (oxygen-rich atmosphere) in a round-bottom flask, leading to the instant formation of a black precipitate. The solution was then refluxed for 30 min at 100~$^\circ$C, and all chemical species dissolved above T~$\simeq$~95~$^\circ$C. Upon gentle cooling, a black solid formed at T = 90~$^\circ$C. At room temperature, the final precipitate was filtered and dried on a sintered glass filter after washing with water and ethanol.\\

\textbf{Pellets Synthesis} -- The powders of the AuClAu compound were embedded (by applying a pressure of 10 tons using a pellet press) into transparent matrices of CsCl, to prevent any possible halide substitution. The pellets were prepared to achieve a composition of 0.1\% by weight of perovskite powder relative to CsCl. \\

\textbf{Raman Spectroscopy} -- The Raman spectra as well as the PL spectra presented in this report were taken using a Horiba Jobin-Yvon 785 nm laser and a CCD detector, with a c-Si sample as reference to correct any instrument bias. \\

\textbf{Room Temperature UV-Visible measurement} -- Absorption data were determined from transmission and reflection measurements of the pellets, conducted using the integrating sphere of the Agilent Cary 5000 UV-Vis-NIR spectrophotometer. Based on the relationship $A + T + R = 100\%$, absorption is calculated as $A = 100 - T - R$. \\

\textbf{PL/PLE measurement} -- Luminescence measurements were conducted on a home-built setup configured for confocal microscopy, featuring a silicon detector (ACTON, TELEDYNE). For the PLE measurement, we use a pulsed laser with a frequency of 78 MHz, scanning wavelengths from 550 nm to 830 nm with a 10 nm bandwidth. Luminescence spectra are measured at each of these wavelengths, and the area under the curve is calculated. The PLE signal is then obtained by normalizing this area to the acquisition time and the number of incident photons. Here, we made sure that the number of incident photons was the same for all excitation wavelengths. For the temperature-dependent measurements, the sample is placed in a cryostat coupled to a liquid nitrogen dewar. The temperature is varied from 300 K to 77 K in steps of 50 K, with a 10-minute wait at each temperature to ensure the sample reaches the target temperature.

\begin{acknowledgement}
The authors express their gratitude to David Cahen and Sarah Houver for invaluable discussions. P.S. thanks the French Agence Nationale de la Recherche (ANR) for funding under the grant number ANR-17-MPGA0012. G.D. thanks the French Agence Nationale de la Recherche (ANR) for funding under the grant number ANR-22-CE42-0027. We thank the CNRS in the framework of a CNRS-Weizmann collaborative program for granting APR's PhD fellowship.
\end{acknowledgement}

\begin{suppinfo}
The supporting information document is available free of charge. It includes more details on the Raman and absorption spectra fitting procedures, an estimation of the PLQY, time-resolved PL analysis, temperature-dependent X-ray diffraction, details on the PLE and RQY estimation, and Urbach tail fittings.
\end{suppinfo}

\bibliography{biblio2}
\end{document}